\documentclass[prl,twocolumn,superscriptaddress,aps,showpacs,amsmath,amssymb,nofootinbib]{revtex4}
\usepackage{graphicx}% Include figure files
\usepackage{dcolumn}%Align table columns on decimal point
\usepackage{bm}% bold math

\begin{document}

\title{Josephson oscillations in binary mixtures of $F=1$ spinor BECs}
\date{\today}
\author{B. Juli\'a-D\'\i az}

\affiliation{Departament d'Estructura i Constituents de la
Mat\`{e}ria, \\
 Universitat de Barcelona, 08028 Barcelona, Spain}
\author{M. Guilleumas}
\affiliation{Departament d'Estructura i Constituents de la
Mat\`{e}ria, \\
 Universitat de Barcelona, 08028 Barcelona, Spain}

\author{M. Lewenstein}
\affiliation{ICREA-Instituci\'o Catalana de Recerca i Estudis Avan\c{c}ats, Llu\'{i}s Companys 23,
08010 Barcelona, Spain}
\affiliation{ICFO-Institut de Ci\`{e}ncies Fot\`{o}niques, 08860 Castelldefels (Barcelona), Spain}

\author{A. Polls}
\affiliation{Departament d'Estructura i Constituents de la
Mat\`{e}ria, \\
Universitat de Barcelona, 08028 Barcelona, Spain}

\author{A. Sanpera}
\affiliation{ICREA-Instituci\'o Catalana de Recerca i Estudis Avan\c{c}ats, Llu\'{i}s Companys 23,
08010 Barcelona, Spain}
\affiliation{Grup de F\'{i}sica Te\`orica, Universitat
Aut\`onoma de Barcelona, 08193 Bellaterra, Spain}

\begin{abstract}

We analyze theoretically Josephson oscillations in a mixture of
two Zeeman states of a spinor Bose-Einstein condensate in a
double-well potential. We find that in the strongly polarized
case, the less populated component exhibits a complex dynamics
with an anti-Josephson behavior, i.e. oscillates in phase with the
more populated one. In the balanced population case the Josephson
oscillations unveal a dependence with the different spin collision
channels. This effect could be used to experimentally measure the
distinct scattering lengths entering in the description of a
spinor condensate. Our numerical results are in close agreement
with an analytical description of the binary mixture using a two
mode model.
\end{abstract}


\maketitle
Tunneling is one of the most fascinating quantum phenomena
known since the development of quantum mechanics. At macroscopic
scale the Josephson effect essentially consists on the fast 
oscillating tunneling through a macroscopic barrier driven by a
quantum phase difference between the two sides of the potential
barrier~\cite{leggett}.

Josephson dynamics of individual atoms has been theoretically
studied by several groups~\cite{tunnellingatoms}. Its extension 
to the tunneling of a Bose-Einstein Condensate (BEC) in a two-well 
potential was presented in~\cite{Smerzi97}. On the experimental 
side, Josephson tunneling of BEC in an optical lattice was first 
reported in Ref.~\cite{kasevich}. Recently, a clear evidence 
of a bosonic Josephson junction in a weakly linked scalar BEC 
has been presented~\cite{Albiez05}. There, the observed time 
evolution of both, the phase difference between the two sides of 
the barrier and the population imbalance shows the importance of 
the interactions and is accurately described by directly solving 
the full Gross-Pitaevskii equation (GP) or by using a two-mode 
approximation~\cite{Gati2007}.

Multi-component BECs trapped in double wells offer an interesting
extension of the tunneling problem. Binary mixtures (i.e.~pseudo-spin 1/2 BECs)
that support density-density interactions have been analyzed in
Refs.~\cite{2component,xu2008}, using mainly the two-mode
approach. More complex spinor BECs, that support population
transfer between different Zeeman states have been also studied as
possible atomic candidates for the macroscopic quantum tunneling
of magnetization (MQTM)~\cite{spin,book}.

In this letter, we investigate the dynamics of a weakly linked
mixture of $m=\pm 1$ Zeeman components of a $F=1$ spinor $^{87}$Rb
condensate. By solving the GP equation, we find that in the strongly 
polarized case the less populated component oscillates in phase with 
the more populated one. For equal populations, the corresponding 
Josephson frequency provides information about the different
scattering lengths of the system. We show how these two combined 
setups give access to the spin dependent 
interaction strenght. Our method extracts information from the 
macroscopic dynamics of the two Zeeman components on a two-well 
potential. It can be seen as the macroscopic counterpart of the 
method of Ref.~\cite{bloch} where the spin dynamics 
of many atom pairs is considered based on an earlier observation 
of coherent spin-dynamics of ultracold atom pairs trapped in sites of an 
optical lattice in a Mott insulator regime. Our method would improve 
the current knowledge, which despite the accuracy of their involved 
experiments 
cannot, for instance, uniquely determine whether the $F=2$ ground 
state of $^{87}$Rb is in an anti-ferromagnetic or a cyclic 
phase~\cite{bloch}. 

Following the conditions used in the experiments of Josephson
tunneling in a scalar condensate~\cite{Albiez05},
we consider $N = 1150$ atoms of spin-1 $^{87}$Rb trapped initially
in the following potential:
\begin{equation}
V({\bf r})=(M/ 2)( \omega_x^2 x^2  +\omega_y^2 y^2 +
\omega_z^2 z^2) +V_0 \cos^2(\pi (x-  \Delta_m)/q_0)\,, \nonumber
\end{equation}
with $\omega_x=2 \pi \times 78$ Hz, $\omega_y=2 \pi \times 66$ Hz,
$\omega_z=2 \pi \times 90$~Hz, $q_0=5.2\, \mu$m, $V_0= 413\, h $
Hz, and $M$ is the mass of the atom. $\Delta_m $ defines the
initial asymmetry in the double-well potential and thus is directly related
to the initial populations imbalance. At $t=0$ the small asymmetry is
switched off ($\Delta_m =0$) and the system is let to evolve in the
double-well potential, which is the same for all components.
The fact that the double well
is created only along the $x$-direction forces the dynamics of the
 Josephson oscillation to be one dimensional
as beautifully demonstrated in~\cite{Albiez05}.
Then, defining $\omega_\perp=\sqrt{\omega_z \omega_y}$
the coupling constants can be rescaled by a factor $1/(2\pi
a_{\perp}^2)$, with $a_{\perp}$ the transverse oscillator length
and the dynamical equations transform into one-dimensional ones~\cite{Moreno2007}.

In the mean-field approach, the $F=1$ spinor condensate is
described by a vector order parameter $\Psi$ whose components
$\psi_m =  |\psi_m| e^{i  \varphi_m}$ correspond to the wave
function of each magnetic sublevel $|F=1, m \rangle \equiv
|m\rangle$ with $m=1,0,-1$. In absence of an external magnetic
field and at zero temperature the spin dynamics of this system
confined in an external potential, $V$, is described by the
following coupled equations for the spin
components~\cite{spinorBEC}:
 \begin{eqnarray}
i \hbar {\partial \psi_{\pm 1}\over \partial t} &=&
 [{\cal H}_s  + c_2(n_{\pm 1}+n_0- n_{\mp 1})]  \psi_{\pm 1}  
+ c_2  \psi_0^2 \psi^{*}_{\mp 1} \,, \nonumber \\ %\label{dyneqs0} \\
i \hbar  {\partial \psi_0\over \partial t} &=&
 [{\cal H}_s  + c_2(n_{1}+n_{-1})]  \psi_{0} 
+ c_2  2 \psi_{1} \psi_0^* \psi_{-1} \,, \label{dyneqs1}
\end{eqnarray}
with ${\cal H}_s=-\hbar^2/(2M)\, {\bm \nabla}^2 +V+c_0n$
being the spin-independent part of the Hamiltonian.
The density of the $m$-th component is given by
$n_{m}({\bf r})=|\psi_m({\bf r})|^2$, while $n({\bf r})=\sum_m|\psi_m({\bf
  r})|^2$ is the total density normalized to the total number of atoms $N$.
The population of each hyperfine state is $N_m=\int d{\bf r} |\psi_m ({\bf r})|^2$.
The couplings are $c_0=4\pi\hbar^2(a_0+2a_2)/(3M)$ and
$c_2=4\pi\hbar^2(a_2-a_0)/(3M)$, where $a_0$ and $a_2$ are the scattering
lengths describing binary elastic collisions in the channels of total spin 0
and 2, respectively. Their values are $a_0=101.8 a_B$ and $a_2=100.4 a_B
$\cite{vanKempen}.
The interatomic interactions permit the transfer of population between the
different Zeeman components by processes that conserve
the total spin, $|0\rangle + |0\rangle \leftrightarrow |+1\rangle + |-1\rangle$.
Therefore, in a spinor condensate the number of atoms
in each component is not a conserved quantity. Due to the chosen conditions,
notably the small total number of atoms and the fact that we will consider
an initial zero population of $m=0$ atoms, the population transfer effects
between the different components is very small \cite{Moreno2007}.
Therefore, in our calculation
the number of atoms in each Zeeman sublevel remains constant in time
as in a real binary mixture, but the spinor character is preserved 
through the parameter $c_2$ which is the spin-dependent collision term.

\begin{figure}[t]
\includegraphics[width=0.75\columnwidth,angle=0, clip=true]{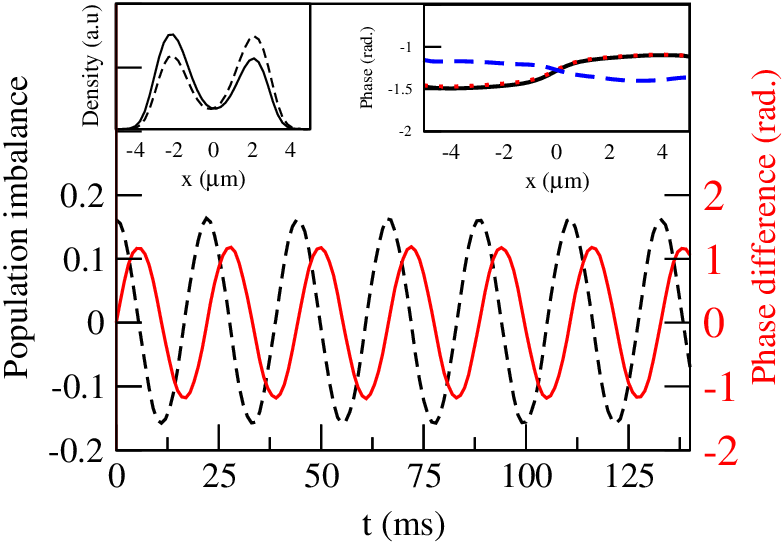}
\caption[]{
Time evolution of the population imbalance (dashed)
and phase difference (solid) for a scalar BEC.
The upper left inset depicts
a snapshot of the population density at t=0 ms (solid) and $t=10$ ms (dashed).
The upper right inset shows the phase as a function of $x$ at
$t=120$~ms. The solid line stands for the scalar BEC
with initial population imbalance 0.06. The dashed and
dotted lines correspond, respectively, to the $m=+1$ and $m=-1$ components
of a simulation
with $(N_{1}/N,N_0/N,N_{-1}/N)=(5\%,0\%,95\%)$ and $z_{\pm1}(0)=\mp 0.06$.
In all cases the initial phase is set to zero for all $x$.}
\label{fig1}
\end{figure}

In the considered double-well potential we can define the 
population imbalances between the two sides of the trap for each
component as: $z_m(t)= [N_{m,L}(t) - N_{m,R}(t)]/  N_m(t)$
where $N_{m,L(R)}(t)$ corresponds to the population on the
left (right) side of the trap. The other relevant quantity is
the phase difference between both sides of the trap defined as,
$ \hat{\varphi}_m(t)=
\hat{\varphi}_{m,R}(t)-\hat{\varphi}_{m,L}(t)$
for each component.
The GP equation predicts the evolution of
$\varphi_m$ as a function of both $t$ and $x$. Within our parameters, 
the condensates at
each side of the trap remain mostly in a coherent state meaning that
the $x$ dependence of $\varphi_m$ is fairly small, as shown in
Fig.~\ref{fig1} where the condensates in each potential well have 
an almost constant phase. Thus, for simplicity, we define
$\hat{\varphi}_{m,L(R)}$ as the phase of $\psi_m(x_M)$, where
$x_M$ is the position of the maximum density
of each condensate.

For the sake of clarity we start by analyzing as a reference
the scalar Josephson effect.
To this aim, we put all the atoms in a single component and solve
the corresponding time dependent GP equation
equivalent to the scalar case with a non linear term proportional to $c_0+c_2$.
In Fig.~\ref{fig1} we show the time evolution of the population
imbalance and phase difference for
$\Delta =-0.20 \mu$m.
Both quantities oscillate with a frequency $\omega_J$
which slightly differs from the bare Rabi frequency
(noninteracting limit) by a term
proportional to the interactions \cite{Smerzi97}.
Two snapshots of the density profile at
$t=0$ and $t=10$~ms can be seen in the inset.

New interesting phenomena appear when a binary mixture
is considered. The binary mixture is simulated by means of
Eqs.~(\ref{dyneqs1}) with an initial zero amount of
$m=0$ atoms. We have checked however, that all the results presented
in this paper are nevertheless
stable against the presence of a small amount of $m=0$
atoms in the trap ($N_0/N\le 1\%$).
We consider $\Delta_{\pm1} =\mp 0.10\mu$m, which
produces asymmetric initial density profiles for both components
with imbalances of $z_{\pm 1}(0)=\mp 0.06$. The configuration
considered is $N_{-1}/N=95\%$ and $N_{1}/N=5\%$ and no initial
phase difference.

\begin{figure}[t]
\includegraphics[width=0.85\columnwidth,
angle=0, clip=true]{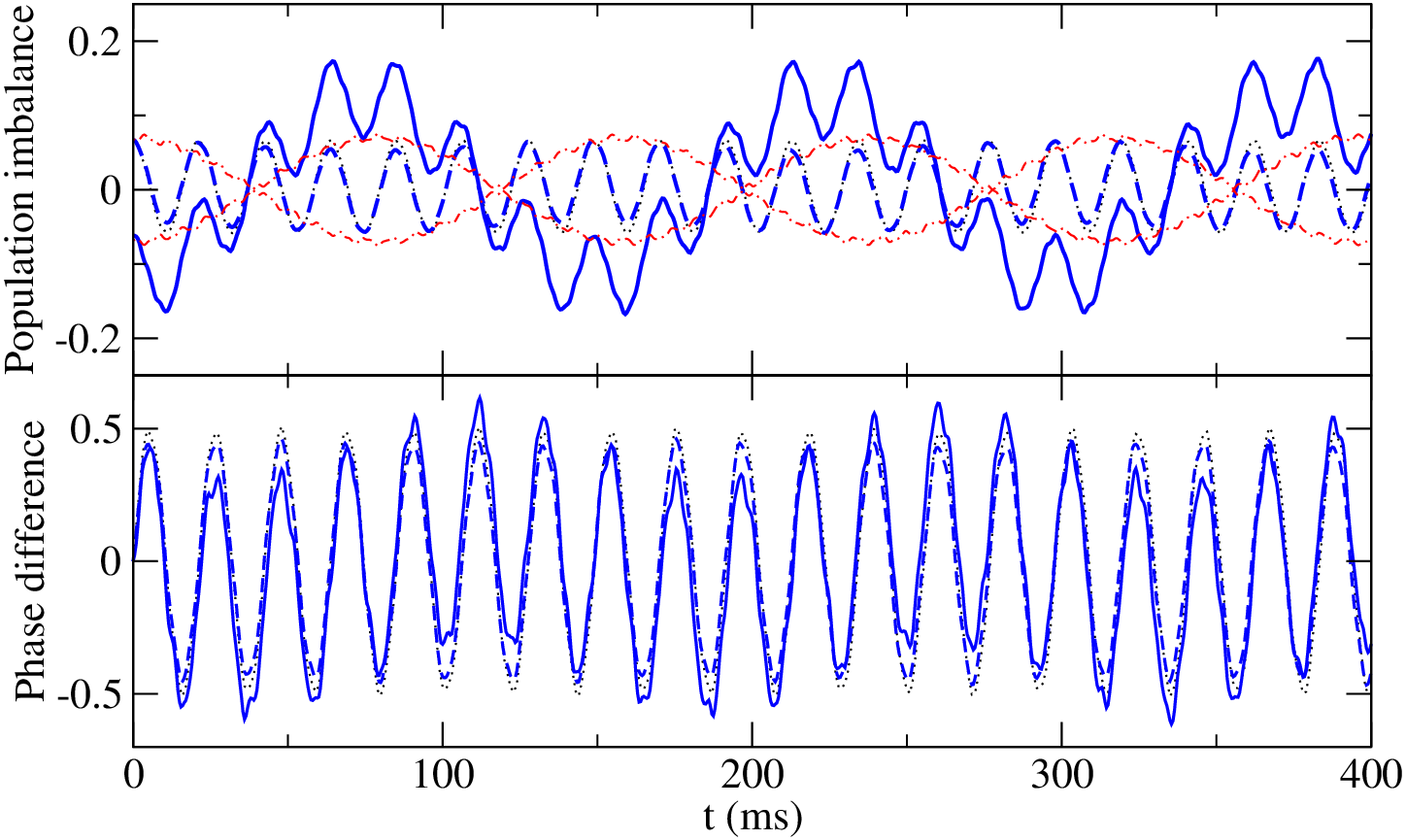}
\caption[]{Population imbalance (upper panel) and phase 
difference (lower panel) between the two sides of each 
condensate. The dotted (black) line corresponds to the 
scalar case, the solid and dashed blue lines correspond to the $m=+1$ 
and $m=-1$ components of a simulation with $(5 \%,0\%,95\%)$. 
The red dot-dashed lines correspond to the $(50\%,0\%,50\%)$ case.   }
\label{fig2}
\end{figure}

The system evolves in the following way. The most populated component
essentially drives the dynamics, its population imbalance and phase
difference, see Fig.~\ref{fig2}, follow a behavior qualitatively and
quantitatively similar to that of the scalar condensate. 
Whereas the less populated
component, $m=+1$ in this case, follows the dynamics of the other one.
First it does an ``anti-Josephson'' tunneling, i.e. the absolute
value of its population imbalance actually grows driven by the other
component, contrary to what would have happened in the absence of the
other component. This is essentially due to the prevalence of the
interaction with the other component over the population
imbalance driven Josephson tunneling.
In addition to the high frequency Josephson oscillation, $\omega_J$, the less
populated component has an additional lower frequency oscillation with a period
of nearly 150 ms. Both effects will be analyzed below within the two mode
approximation.

\begin{figure}[t]
\includegraphics[width=0.65\columnwidth,
angle=0, clip=true]{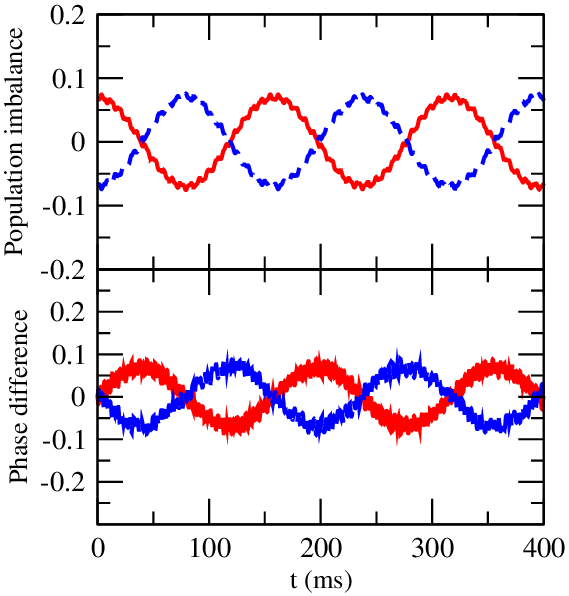}
\caption[]{Population imbalance (upper panel) and phase difference
(lower panel) between the two sides of each BEC for a simulation
with $(50\%, 0\%, 50\%)$. Solid (red) lines correspond to the $m=-1$ atoms
and dashed (blue) to the m=+1 atoms.}
\label{fig3}
\end{figure}

A second interesting configuration is that of having the same
population for both $m=\pm 1$ components, $(50\%, 0\%, 50\%)$ with the
same initial population imbalance as before, $z_{\pm 1}(0)=\mp 0.06$. In
this case the behavior is completely different. Essentially, as can
be seen in Fig.~\ref{fig3}, a Josephson tunneling with an oscillation
period of approximately 150 ms (almost 6 times larger than in the scalar case)
is observed for both components. Thus, the longer
oscillation, which actually corresponds to the lower frequency oscillation
seen in Fig.~\ref{fig2}, is a pure effect of having a binary mixture.

To get a deeper insight on the observed behavior we have performed a
two mode analysis of a binary mixture of BECs.
The general coupled GP equations for a binary mixture of components
$a$ and $b$, which keeps invariant the total number of atoms
of each species, can be written as:
\begin{eqnarray}
i\hbar{\partial \Psi_a\over \partial t} &=& \!\!\!
\left[-{\hbar^2 \nabla^2 \over 2m_a}
+V_a +g_{aa}|\Psi_a|^2 +g_{ab}|\Psi_b|^2 \right]\Psi_a\nonumber \\
i\hbar{\partial \Psi_b\over \partial t} &=& \!\!\!
\left[-{\hbar^2\nabla^2 \over 2m_b}
+V_b +g_{bb}|\Psi_b|^2 +g_{ba}|\Psi_a|^2\right] \Psi_b\,. \label{eq:mgp}
\end{eqnarray}
These general couplings, $g_{ij}$, can be related to the $c_0$ and $c_2$ ones
entering in Eqs.~(\ref{dyneqs1}) as $g_{aa}=g_{bb}=c_0+c_2$
and $g_{ab}=g_{ba}=c_0-c_2$. Since for $^{87}$Rb atoms in $F=1$ 
$|c_2|<< c_0$, the mixture we are considering
has the same self interaction between each
component and similar to the mixing term, $g_{ab}\sim g_{aa}$.

The two mode approximation can be expressed with the
following ansatz \cite{Smerzi97}:
$\Psi_j= \Psi_{j L}(t)\Phi_{j L}(r) + \Psi_{j
  R}(t)\Phi_{j R}(r)$, with
$\langle \Phi_{i\alpha}|\Phi_{j\beta}\rangle=\delta_{ij}\delta_{\alpha\beta}$ and
$\Psi_{j,\alpha}(t)=\sqrt{N_{j,\alpha}(t)} e^{\imath \phi_{j,\alpha}(t)}$,
$i,j=a,b$ and $\alpha,\beta=L,R$.
Equations for the coefficients $\Psi_{i,\alpha}(t)$ are obtained from Eqs.~(\ref{eq:mgp})
neglecting terms involving mixed products of
$L$ and $R$ wave functions of order larger than 1.  Rewritten in terms of the imbalances
$z_j(t)= [N_{j,{\rm L}}(t) - N_{j,{\rm R}}(t)]/  N_j(t)$, and phase differences,
$\phi_j=\phi_{j,R}-\phi_{j,L}$ they take the form
\begin{eqnarray}
\dot{z_a}&=& - \omega_r \sqrt{1-z_a^2}\sin {\phi_a}
\;,\,
\dot{z_b}= - \omega_r \sqrt{1-z_b^2}\sin {\phi_b}
  \nonumber \\
\dot{\phi_a}&=& { U\over \hbar} N_a z_a +{\tilde{U} \over  \hbar}  N_b z_b
+{\omega_r z_a\over \sqrt{1-z_a^2}}\cos \phi_a\nonumber \\
\dot{\phi_b}&=& {U \over \hbar} N_b z_b +  {\tilde{U}\over \hbar}N_a z_a
+{\omega_r z_b\over \sqrt{1-z_b^2}}\cos \phi_b\,,
\label{eq:mgs6-simp5}
\end{eqnarray}
where $\omega_r=2 K/\hbar$ denotes the Rabi frequency, 
and
\begin{eqnarray}
K&=&-\int d\vec{r} \bigg[
{\hbar^2\over  2M}  \,\nabla\Phi_{L}(\vec{r})\cdot \nabla\Phi_{R}(\vec{r})
+\Phi_{L}(\vec{r}) \,V\,\Phi_{R}(\vec{r})\bigg]
\nonumber \\
U_{ij}&=&g_{ij}\int d\vec{r} \Phi_{i,L}^2(\vec{r})\Phi_{j,L}^2(\vec{r})
     =g_{ij}\int d\vec{r} \Phi_{i,R}^2(\vec{r})\Phi_{j,R}^2(\vec{r})\,.
     \nonumber 
\end{eqnarray}
In Eqs. (\ref{eq:mgs6-simp5}) we have considered a
symmetric double-well and the same mass for the particles 
of both components $m_a=m_b \equiv M$, thus defining 
$\Phi_{L(R)}\equiv \Phi_{a,L(R)}=\Phi_{b,L(R)}$. This yields:
$U_{aa}=U_{bb}\equiv U$ and
$U_{ab}=U_{ba}\equiv \tilde{U}$.
The stability of these equations has
been analyzed in Ref.~\cite{xu2008}.

The dynamics obtained by solving the coupled GP
Eqs.~(\ref{dyneqs1})
can be now understood by taking the appropriate
limits on Eqs.~(\ref{eq:mgs6-simp5}), after linearizing them around the
equilibrium values
$z_a=z_b=0=\phi_a=\phi_b$. Defining
$\alpha$ and $\beta$
as $N_a=N (1-\alpha)$, $N_b=N\alpha$ and $\tilde{U}=U(1+\beta)$,
Eqs.~(\ref{eq:mgs6-simp5}) reduce to
\begin{eqnarray}
\dot{z_a}&=&-\omega_r \phi_a \;,\quad
\dot{z_b}= -\omega_r \phi_b \nonumber \\
\dot{\phi_a}&=& \left[(1-\alpha) NU/\hbar +\omega_r\right] z_a 
+ \alpha (1+\beta)NU/\hbar \;z_b \nonumber \\
\dot{\phi_b}&=& (1-\alpha)(1+\beta) NU/\hbar \;z_a +
\left[\alpha NU/\hbar + \omega_r\right] z_b \,.
\label{eq:linear}
\end{eqnarray}
These equations correspond to two coupled nonrigid pendulums 
with normal mode frequencies $\omega_>$ and $\omega_<$.
In our system,  $g_{ab}\sim g_{aa}=g_{bb}$, i.e. $\tilde{U}\sim U$
and thus $\beta<<1$. Then the eigenfrequencies can be approximated by:
$\omega_>= \omega_r + {NU\over \hbar}\alpha\beta (\alpha-1)$, 
$\omega_<=
\omega_r {\left[1 + {NU\over \hbar \omega_r}(1+\alpha \beta-\alpha^2\beta)\right]\over
\sqrt{1+{NU \over \hbar \omega_r}}}$.
The behavior observed in Fig.~\ref{fig2} is explained by taking
$\alpha\!\to\! 0$.
 In this limit
the frequencies simplify to $\omega_>=\omega_r$ and
$\omega_< = \omega_r\sqrt{1+UN/(\hbar \omega_r)}=\omega_J$ .
The highest populated component, $z_a$, decouples from the less populated
and thus performs a Josephson oscillation,
as shown in Fig.~\ref{fig2}, with the same frequency as the scalar case
$\omega_J$.
And the less populated component oscillates with the two 
eigenfrequencies:
$\omega_>=\omega_r$ and $\omega_< =\omega_J$.

The case of equal populations can also be addressed within
Eqs.~(\ref{eq:linear}) by setting $\alpha=1/2$. In this case the
long mode can be enhanced by starting with opposite initial imbalances
as in Fig.~\ref{fig3}.
Then, both components oscillate with
$\omega_>=\omega_r - {NU\over 4\hbar}\beta\equiv \tilde{\omega}$. 
Since $\beta << 1$,
this explains the similarity between the oscillation frequency
of the $(50\%,0\%,50\%)$ case studied and the long oscillation of the 
less populated component of the $(5\%,0\%,95\%)$ case,
see Fig.~\ref{fig2}. Let us emphasize that the frequency shift with 
respect to the Rabi frequency, $NU/ (4\hbar)\beta$, is a direct 
consequence of the difference between
$g_{aa}$ and $g_{ab}$, which in turn depend on the interaction
strengths, $c_0$ and $c_2$.
Alternatively, if the scattering lengths are known, the Josephson 
oscillation provides a tool to determine
with high precision the number of atoms.

It is important to mention that
we have assumed zero temperature and therefore, no phase fluctuations
are present in our calculation. Indeed,
phase fluctuations will evidently destroy the phase coherence of
the condensate in each potential well leading to a
complete suppression of Josephson oscillations.
For very elongated systems, it is known that even at small
temperatures the phase
of the condensate fluctuates~\cite{Petrov}.
However, in the setup we have considered, which is closely
related to the experimental setup demonstrating Josephson oscillations in
a scalar BEC~\cite{Albiez05},
phase fluctuations due to thermal effects will remain small since
the condensate is essentially three dimensional
although the dynamics is one dimensional. A detailed analysis of possible 
noise sources using a two mode approximation has been developed for a 
scalar condensate in a double well
potential in~\cite{Gati2007}, showing that coherence is preserved at the 
temperatures at which the experiment runs, and demonstrating
also how coherence is lost by increasing the temperature.
A similar analysis applies here with a spinor BEC.

Summarizing, we have addressed the problem of a binary mixture of
BECs which is realized with a $F=1$ spinor BEC where the $m=0$
component is not populated. The behavior of the mixture has been
analyzed in two significant limits: a binary mixture with a large
difference in the concentration of the two components
and a binary mixture with equal amount of both components. Both cases have been
studied with the GP equations and have also been scrutinized by
employing a simple two mode approximation.

An independent measurement of the Rabi frequency could be obtained from the long 
oscillation of the less populated component of Fig.~\ref{fig2}. That 
would allow for precise determination of the (otherwise difficult to 
measure, cf. \cite{bloch}) spin-dependent interaction strength, $a_2-a_0$,
using the relations, $(a_2-a_0)/(a_0+2a_2)= -\beta/(2+\beta)$ and
$\beta$ can be evaluated by using the above frequencies as: 
$\beta=4\omega_r(\omega_r- \tilde{\omega})/(\omega_J^2-\omega_r^2)$.  
The accuracy of the method is expected to be comparable to that of
Ref.~\cite{bloch} and is only constrained by the ability to measure 
with precision the different macroscopic oscillations. The method presented 
could be extended to any atom species and would be easy to generalize to the $F=2$ case. 

We thank M. Oberthaler for encouraging us to develop this research direction.
B.J-D. is supported by a CPAN CSD 2007-0042 contract.
This work is also supported by
Grants No. FIS2008-01661, FIS2008-00421, FIS2008-00784, FerMix,  and QOIT from MEC.

Note added. After the submission of this paper, a
related study appeared, reporting a two-mode 
study of binary mixtures~\cite{last}.

\end{document}